\documentclass[a4paper,twoside]{article}
\usepackage{AMS2LA,fancyhdr,CJK,multicol,graphicx,bm,ccmap,float}
\def\headrule{\kern 1mm \hrule width 17cm \kern -1mm}%
\def\footnoterule{\kern 1mm \hrule width 7cm \kern 2.2mm}%
\def\REF#1{\par\hangindent\parindent\indent\llap{#1\enspace}\ignorespaces}%

\newcommand{\cplyear}{2015} \newcommand{\cplvol}{XX}
\newcommand{\cplno}{X} \newcommand{\cplpagenumber}{XX{XXX}}
 \newcommand{\cplpage}{\cplpagenumber-\thepage}
\begin{document} \begin{CJK}{UTF8}{gbsn}\vspace* {-6mm} \begin{center}
\large\bf{\boldmath{Optimal Size for Maximal Energy Efficiency in Information Processing of Biological Systems Due to Bistability}}
\footnote{This work was supported by the National Natural Science Foundation
of China (Grants No. 11105062), the Fundamental Research Funds for
the Central Universities (Grant No. lzujbky-2011-57 and No. lzujbky-2015-119).

\hspace*{1.8mm}$^{**}$Correspondence author. Email:
yulch@lzu.edu.cn

\hspace*{1.8mm}\copyright\,{\cplyear}
\href{http://www.cps-net.org.cn}{Chinese Physical Society} and
\href{http://www.iop.org}{IOP Publishing Ltd}}
\\[4mm]
\normalsize \rm{} ZHANG Chi(张弛)$^{1}$, LIU Li-Wei(刘利伟)$^{2}$，WANG Long-Fei(王龙飞)$^{3}$， YUE Yuan(岳园)$^{3}$，YU Lian-Chun(俞连春)$^{3,4**}$
\\[1mm]\small\sl $^{1}$Cuiying Honors College, lanzhou University, Lanzhou 730000

$^{2}$ College of Electrical Engineering, Northwest University for Nationalities, Lanzhou 730070, China

$^{3}$ Institute of Theoretical Physics, Lanzhou University,Lanzhou 730000, China

$^{4}$ Key Laboratory for Magnetism and Magnetic Materials of the Ministry of Education, Lanzhou University, Lanzhou 730000, China
\\[4mm]\normalsize\rm{}(Received XXX)
\end{center}
\end{CJK}
\vskip -1mm

\noindent{\narrower\small\sl{}Energy efficiency is closely related to the evolution of biological systems and is important to their information processing.
In this paper, we calculated the excitation probability of a simple model of a bistable biological unit in response to pulsatile inputs, and its spontaneous excitation rate
due to noise perturbation. Then we analytically calculated the mutual information, energy cost, and energy efficiency of an array of these bistable units.
We found that the optimal number of units could maximize this array's energy efficiency in encoding pulse inputs, which depends on the fixed energy cost.
We conclude that demand for energy efficiency in biological systems may strongly influence the size of these systems under the pressure of natural selection.

\par}\vskip 3mm\normalsize

\noindent{\narrower\sl{PACS:05.10.Gg, 05.40.Ca, 02.50.-r}
{\rm\hspace*{13mm}DOI: XXX }

\par}\vskip 3mm
\begin{multicols}{2}

The use of the ideas of statistical mechanics to study biological systems are nearly as old as this disciplines itself.
Recently, statistical physics has been demonstrated as a useful tool to understand the energy efficiency of biological systems, which measures their energy cost for performing specific functions,
from sensory adaptation in biochemical negative feedback loops to information processing in the brain. $^{[1, 2]}$ It is believed that energy efficiency plays an
important role in guiding the directions to which the physiology and anatomy of biological systems have evolved, and it may be related
to the basic laws of statistical physics in a fundamental way. $^{[1]}$

There are many examples of biological systems which can operate, in a stable manner, in two distinct modes.  This bistability is a typical way for a biological system
to transmit information digitally.$^{[3]}$ For instance, the initiation of action potentials in neurons follows an ``all or none'' principle,
and an action potential is generated only when the system crosses a certain threshold.$^{[4]}$  Further examples can be found in subcellular signal transduction networks.$^{[5]}$
Since biological systems are generally subjected to noise, the signal processing within a single pathway is potentially unreliable.$^{[6,7]}$ Thus, the multiple pathway
is a reasonable choice to out of this predicament.$^{[8]}$ For example, information process in the brain is accomplished by a group of neurons working cooperatively,
 and studies have shown that neurons may become synchronized to minimize perturbation from noise and thereby facilitate reliable information transmission.$^{[9]}$

Although transducing information through multiple pathways can improve reliability in the face of noise, it can cause a burdensome energy cost to the system.
For example, the human brain accounts for about 20\% of an adult's resting metabolic rate, and a large fraction of this energy is used for the generation
of action potentials.$^{[10],[11]}$ Therefore, metabolic demands can be large enough to influence the design, function, and evolution of these biological systems
under the pressure of natural selection through evolution,$^{[12,13]}$ so that they might be more energy efficient.$^{[14-15]}$

Studies have shown that energy efficiency in biological systems is greatly influenced by physical size. Taking neural systems as an example, smaller neurons will
cost less energy because fewer ion channels are involved and there is less ion exchange through ion pumps that drive ATPase Na+/K+ exchangers in the recovery from an action potential.$^{[16]}$
     However, the stochastic nature of ion channel gating creates variability in the response of neuron to external stimuli, and spontaneous action potentials,
      which reduce the reliability of signal processing.$^{[17]}$ In this case, trade-offs between information transfer and energy use could strongly influence the number of ion channels used by the neurons.
       Recent theoretical analysis and computer simulations showed that the energy efficiency could be maximized by the number of ion channels in a single neuron, or the number of neurons in a network.$^{[18],[19]}$

   Since bistability is a general mechanism for biological information processing, it is interesting to study the dependence of energy efficiency on system size in this context.
    In this paper, we first solve a one-dimensional bistable Langevin equation, which mimics the switching of a biological system unit between a resting and excited state using a particle crossing the barrier of a double well potential.
    Using an analytical solution for the pulse signal detection rate and spontaneous excitation rate, we theoretically calculate the mutual information and energy cost of a
    biological unit array to measure its energy efficiency. Lastly, we find an optimal number of biological units in this array that maximize its energy efficiency.

The dynamics of our biological model unit is described with the following equation:
\begin{equation}\label{eq1}
\dot x =  - U'(x) + \Gamma (t),
\end{equation}
where $x$ is the unit's physiological variable, for example, the membrane potential of a neuron. $U$ is a double well potential, defined as:
\begin{equation}\label{eq2}
U =  - \frac{a}{2}x^2  + \frac{{x^4 }}{4}.
\end{equation}
$U$ has two minima $x_{s1}  =  - \sqrt a$, $x_{s2}  = \sqrt a$ and a saddle point $x_\mu   = 0$. $\Gamma (t)$ is a Gaussian random variable:
 \begin{equation}\label{eq3}
 < \Gamma (t) >  = 0;\; < \Gamma (t)\Gamma (t') >  = 2D\delta (t - t'),
 \end{equation}
where D is noise intensity.

\begin{figure}[H]
 \centering
  \includegraphics[scale=0.25]{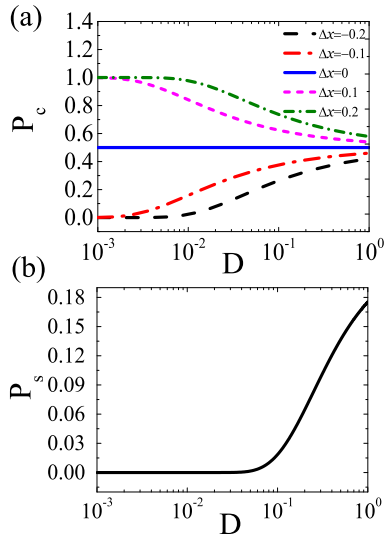}\\
 \caption{The detection rate of the bistable unit as a function of noise intensity for different input pulse strength (a) and its spontaneous excitation rate as a function of noise intensity (b).}\label{f1}
\end{figure}

The unit is at its resting state when the particle is in the left well, and is excited when the particle crosses the barrier to the right well due to noise or an applied signal
perturbation. We assume that the particle is subjected to a force of short duration, which horizontally moves the particle into the region of the saddle point.
After removing the force, the particle drifts up to the region of the saddle point. Near the saddle point, the repelling phase trajectories cause the particle to accelerate away
from the saddle point region towards one of the two minima. Following Lecar and Nossal's approach of linearizing around the saddle point,$^{[20]}$ we obtained the probability
 of finding the particle in the right well after a long enough time.

First, we expand Eq.\,(1) near the threshold singular point. Defining
\begin{equation}\label{eq4}
\delta  = x - x_\mu
\end{equation}
and since $x_\mu= 0$, we obtain the equation:
\begin{equation}\label{eq5}
\dot \delta  = a\delta  + \Gamma (t).
\end{equation}
The solution of Eq.\,(5) is
\begin{equation}\label{eq6}
\delta (t) = \delta (0)e^{at}  + \int_0^t {e^{a(t - s)} \Gamma (s)} ds,
\end{equation}
and the integral equals:
\begin{equation}\label{eq7}
\int_0^t {e^{a(t - s)} \Gamma (s)} ds = \delta (t) - \delta (0)e^{at}  = X(t).
\end{equation}
Since $\Gamma(s)$ is a Gaussian random variable, the time integral $X(t)$ also obeys a Gaussian distribution, so we have
\begin{equation}\label{eq8}
P(X,t) = (2\pi  < X^2  > )^{ - \frac{1}{2}} \exp (\frac{{ - X^2 }}{{2 < X^2  > }}).
\end{equation}
We express the expectation value of $X^2$ in terms of the joint expectation of the variable $\Gamma$  taken with itself at a different time. From Eq.\,(7) we have
\begin{equation}\label{eq9}
 < X^2  >  = \frac{D}{a}(e^{2at}  - 1).
\end{equation}
The unit's excitation probability under a pulse input is equal to the probability that $\delta (0) > 0$ when $t \to \infty$, given an initial displacement,  $\delta(0)$, then,
\begin{equation}\label{eq10}
\begin{array}{r@{~}l}
P[\delta (t) > 0|\delta (0)] &=\mathop {\lim }\limits_{t \to \infty } \int_{ - \delta (0)e^{at} }^\infty  {P(X,t)dX}\\
                             &=\frac{1}{2}[1 + erf(\frac{{\delta (0)}}{{\sqrt {2D/a} }})],
\end{array}
\end{equation}
where $erf(x)$ is the Gaussian error function, which has the form
\begin{equation}\label{eq11}
erf(x) = \frac{2}{{\sqrt \pi  }}\int_0^x {\exp ( - t^2 )} dt.
\end{equation}
For simplicity, defining $\delta (0) = \Delta x $, where $\Delta x $ is the input strength, we can rewrite the probability that the unit is excited after application of a pulse input (a.k.a. the signal detection rate):
\begin{equation}\label{eq12}
P_c (\Delta x) = \frac{1}{2}[1 + erf(\frac{{\Delta x}}{{\sqrt {2D/a} }})].
\end{equation}

Figure 1(a) shows the probability of the unit being excited as a function of noise intensity $D$ with different input pulse strengths $\Delta x$ described by Eq.\,(12).
The detection rate in response to threshold input($\Delta x=0$) is 0.5 and is independent of noise intensity. The detection rate in response to subthreshold inputs ($\Delta x<0$)
increases as the noise intensity increases. This means that, with the assistance of noise, the bistable unit can detect subthreshold signals, which is a well-known result known as stochastic resonance.
 However, the noise sabotages the neuron's reliability when receiving suprathreshold inputs ($\Delta x>0$) and the detection rate decreases as the noise intensity increases.

In the absence of inputs, the unit can be excited under the perturbation of only noise. This spontaneous excitation rate of this bistable unit can be calculated according to Kramers's formula for escaping rate,$^{[21]}$ which is written as
\begin{equation}\label{eq13}
K = \frac{1}{{2\pi }}\sqrt {U''(x_s1 )|U''(x_u )}| \exp ( - \frac{{\Delta U}}{D}),
\end{equation}
where $\Delta U =U(x_u)- U(x_{s1})$. Thus, the spontaneous excitation rate of the above bistable unit is
\begin{equation}\label{eq14}
P_s  = \frac{{\sqrt 2 a}}{{2\pi }}\exp ( - \frac{{a^2 }}{{4D}}).
\end{equation}
As the noise intensity increases, more threshold-crossing events may happen due to noise, thus the spontaneous excitation rate increases as the noise intensity increases (Fig. 1 (b)).

Next, we consider a bistable unit array model as shown in Fig. 2(a).  In this model, each unit in the array receives identical pulsatile input, and the output of this array is the composition of the outputs of each unit. Assuming the input strength is distributed uniformly over the interval $[\Delta x_{\min } ,\Delta x_{\max } ]$, \emph{i.e.}, its probability distribution follows
\begin{equation}
q(\Delta x) = \frac{1}{{\Delta x_{\max }  - \Delta x_{\min } }}=\frac{1}{\Delta s}.
\end{equation}
Then, $\bar x = \frac{{(\Delta x_{\min }  + \Delta x_{\max } )}}{2}$ is the mean value of input strength. In the following calculation, we fix this distribution interval between $[-0.1, ~0.1]$ so that both the subthreshold and suprathreshold inputs are involved. With the input $\Delta x \in S$, the output of this unit array is discrete, i.e., $R=\{r|r=K / N, K=0,1,2,\ldots,N \}$, where $K$ is the number of units excited after the inputs are applied.

\begin{figure}[H]
 \centering
  \includegraphics[scale=0.25]{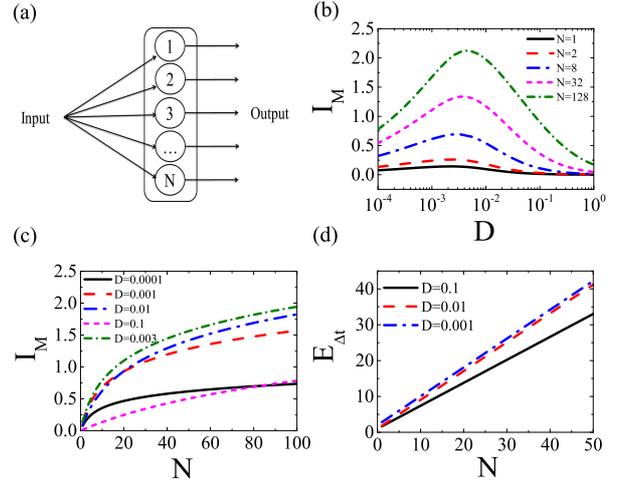}\\
 \caption{(a) The biological array is composed of $N$ independent bistable units; (b) The mutual information depends on noise intensity for different array size; (c) Mutual information as a function of number of units for different noise intensity; (d) Energy cost depends on number of units in the array  for different noise intensity.}\label{f2}
\end{figure}

Since each unit in the array is independent, the conditional probability $q(r|\Delta x)$ that the output is $r$ when the input is $\Delta x$ is given by a binomial distribution:
\begin{equation}
q(r|\Delta x) = \left( {\begin{array}{*{20}c}
   N  \\
   K  \\
\end{array}} \right)(P_c (\Delta x))^K  \cdot (1 - P_c (\Delta x))^{N - K},
\end{equation}
where $
\left( {\begin{array}{*{20}c}
   N  \\
   K  \\
\end{array}} \right)
$ is the binomial coefficient and $ P_c (\Delta x)$ is the probability that a unit is excited after receiving a pulse input.
According to Bayes formula, the probability that the output is $r$ can be obtained by
\begin{equation}
q(r) = \int\limits_S {q(\Delta x) \cdot q(r|\Delta x)} d(\Delta x)= \frac{1}{\Delta s} \int\limits_S q(r|\Delta x) d(\Delta x).
\end{equation}
According to Shannon's information theory,$^{[22]}$ the information between input $S$ and output $R$ is defined as
\begin{equation}
I_M(S;R) = \sum\limits_{\Delta x \in S} {\sum\limits_{r \in R} {q(\Delta x)} } q(r|\Delta x)\log _2 \frac{{q(r|\Delta x)}}{{q(r)}}.
\end{equation}
In our case of unit arrays, the input is continuous and output is discrete, thus the summation need to be rewrite as follows:
\begin{equation}
I_M(S;R) = \sum\limits_{r \in R} {\int\limits_S {q(may cause the page layout to be inconsistent, however.\Delta x)} } q(r|\Delta x)\log _2 \frac{{q(r|\Delta x)}}{{q(r)}}d(\Delta x).
\end{equation}
Finally, we obtain
\begin{equation}
I_M(S;R) = \frac{1}{\Delta s}\sum\limits_{K = 0}^N {\int_{\Delta x_{\min } }^{\Delta x_{\max } } {q(r|\Delta x)} \log _2 \frac{{q(r|\Delta x)}}{{q(r)}}}d(\Delta x)
\end{equation}

 Figs. 2(b) and (c) show the dependence of mutual information on noise intensity and the number of units, respectively.  We note that the mutual information exhibits maximal values for moderate value of noise intensity.
 This phenomenon is known as suprathreshold stochastic resonance, and the same results were reported in Ashida and Kubo's work,$^{[23]}$ in which the detection function was obtained
 by fitting the S-shaped response curves (from simulations of a stochastic channel-based model) to an integrated Gaussian function.

For an array of containing $N$ units, the totally energy expenditure in a time interval $\Delta t$ can be written as
\begin{equation}
E_{\Delta t}  = E_0 \Delta t + E_{n} (N)\Delta t + \int\limits_S {d(\Delta x)q(\Delta x)E_{s} (N,\Delta x)},
\end{equation}
where $E_0$ is the fixed energy cost in unit time, which is independent of the number of units in the array. The last two terms are related to the energy cost due to excitation. For simplicity, we assume the energy cost of one excitation is 1. $E_n(N)$ is the energy cost of the spontaneous excitation due to noise in unit time, namely $E_n(N)=NP_s$. $E_{s} (N,t)$ is the energy cost of the excitation in response to input pulses with strength $\Delta x$  and $E_{s} (N,t) = NP_c (\Delta x)$ if the inputs are applied in this time interval, else its value is zero. Therefore, $\int\limits_S {d(\Delta x)q(\Delta x)E_{s} (N,t)}$  is the average energy cost of excitations in response to input pulses with distribution $q(\Delta x)$. Fig. 2(d) shows the dependence of energy cost on the number of units in the array.

Now we consider the energy efficiency of this bistable array. After inputs are turned on, the mutual information in unit time is $I = \frac{{I_M }}{{\Delta t}}$ and the energy cost in unit time is $E = \frac{{E_{\Delta t} }}{{\Delta t}}$. Therefore, we define energy efficiency as
\begin{equation}
\eta  = \frac{I}{E} = \frac{{I_M }}{{E_0 \Delta t + NP_s \Delta t + \int\limits_S {d(\Delta x)q(\Delta x)NP_c (\Delta x)} }},
\end{equation}
which measures how many bits of information are encoded by the systems with the consumption of one unit of energy.

\begin{figure}[H]
 \centering
  \includegraphics[scale=0.25]{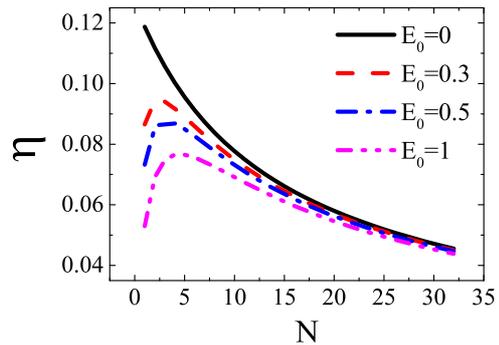}\\
 \caption{The energy efficiency of the biological unit array as a function of unit number for different fixed energy $E_0$. $\bar v=0.1$.}\label{f1}
\end{figure}

Fig. 3 shows the energy efficiency of the above biological array as a function of its size. It is seen that as the number of units increases, the energy efficiency first increases and then drops, so that a maximum exists. As the fixed energy cost $E_{0}$ increases, the optimal size decreases. When $E_{0}=0$, the maximum disappears because the energy efficiency decreases monotonously as the array size increases.

We see that the fixed energy cost is vital for the energy efficiency being maximized by the array size.
 Indeed, Schreiber \emph{et al.}, studied the energy efficiency of a group of ion channels transmitting information with graded electrical signals.
 They found that an optimum number of channels could maximize energy efficiency. The optima depend on the fixed energy cost that is related to costs that have to be met in the absence of signals, such as the synthesis of proteins and lipids.$^{[18]}$ Here, in our work, the fixed energy cost is essential for the energy efficiency to have a maximum. We argue that this fixed energy cost is independent of the array size. For instance, this cost could correspond to the energy that downstream
  systems expend to read out the information in the synchronized excitation events, as our previous work on the neural systems has showed.$^{[7]}$

In conclusion, we have analyzed the energy efficiency of an array of biological units that function according to a simple bistable model.
We demonstrated that the biological array exhibits maximal energy efficiency with an optimal number of units in the array. We interpret this
to imply that the demands placed on biological systems for energy efficiency under the pressure of natural selection could be strongly influenced by their size.
This conclusion is consistent with previous studies on single neurons and neuronal populations.$^{[18,19]}$ Since a bistable model is also often used in physical systems
 for communication systems, the principles we have demonstrated here suggest a possible way to build more energy efficient communication devices for noisy environments.

\section*{\Large\bf References}%

\vspace*{-0.8\baselineskip}

\hskip 7pt

{\footnotesize%

\REF{[1]} Lan G H, et al 2012 {\it Nat. Phys.} {\bf 8} 422
\REF{[2]} Sengupta B, Stemmler M B and Friston K J 2013 {\it PLoS Comput. Biol.} {\bf 9} e1003157
\REF{[3]} Pomerening J R 2008 {\it Curr. Opin. Biotechnol.} {\bf 19} 381
\REF{[4]} Hodgkin A L and Huxley A F 1952 {\it J. Physiol.} {\bf 117} 500
\REF{[5]} Ma L et al 2005 {\it PNAS} {\bf 102} 14266
\REF{[6]} Pilpel Y 2011 Methods Mol Biol \textbf{759} 407
\REF{[7]} Chen Y, Yu L C and Qin S M 2008 {\it Phys. Rev. E.} {\bf 78} 051909
\REF{[8]} Simpson M L et al 2009 {\it Wiley Interdiscip. Rev. Nanomed. Nanobiotechnol.} {\bf 1} 214
\REF{[9]} Tabareau N, Slotine J J, and Pham Q C 2010 {\it PLoS Comput. Biol.} {\bf 6} e1000637
\REF{[10]} Richter D 1957 Metabolism of the nervous system. (New York: Elsevier)
\REF{[11]} Attwell D and Laughlin S B 2001 {\it J. Cereb. Blood Flow Metab.} {\bf 21} 1133
\REF{[12]} Laughlin S B, van Steveninck R R D R and Anderson J C 1998 {\it Nature neurosci.} {\bf 1} 36
\REF{[13]} Yu Y, Hill A P and McCormick D A 2012 {\it PLoS Comput. Biol.} {\bf 8} e1002456
\REF{[14]} Niven J E and Laughlin S B 2008 {\it J. Exp. Biol.} {\bf 211} 1792
\REF{[15]} Laughlin S B 2001 {\it Curr. Opin. Biotechnol.} {\bf 11} 475
\REF{[16]} Niven J E and Farris S M 2012 {\it Curr. Biol.} {\bf 22} R323

\REF{[17]} Chow C C and White J A 1996 {\it Biophys. J.} {\bf 71} 3013
\REF{[18]} Schreiber S el at 2002 {\it Neural Comput.} {\bf 14} 1323
\REF{[19]} Yu L and Liu L 2014 {\it Phys. Rev. E.} {\bf 89} 032725
\REF{[20]} Lecar H and Nossal R 1971 {\it Biophys J.} {\bf 11} 1048
\REF{[21]} Gardiner C 2009 \emph{Stochastic Methods: A Handbook for the Natural and Social Sciences} (Berlin: Springer)
\REF{[22]} Shannon C E and Weaver W 1948 \emph{The mathematical theory of communication.} (Urbana: University of Illinois Press)
\REF{[23]} Ashida G and Kubo M 2010 {\it Physica D: Nonlinear Phenomena.} {\bf 239} 327

}
\end{multicols}
\end{document}